\documentclass[fleqn,10pt]{wlscirep}
\usepackage[utf8]{inputenc}
\usepackage[T1]{fontenc}

\usepackage{amssymb}
\usepackage{amsmath}
\usepackage{hyperref}

\title{The explosive value of the networks }

\author[1,2,*]{Antonio Scala}
\author[2,+]{Marco Delmastro}
\affil[1]{CNR-ISC, Applico Lab, Roma, 00185, Italy}
\affil[2]{Centro Ricerche Enrico Fermi, Roma, 00184, Italy}

\affil[*]{antonio.scala@cnr.it}
\affil[+]{marco.delmastro@cref.it}

\begin{abstract}
Networks have always played a special role for human beings in shaping social relations, forming public opinion, and driving economic equilibria. Nowadays, online networked platforms dominate digital markets and capitalization leader-boards, while social networks drive public discussion. Despite the importance of networks in many economic and social domains (economics, sociology, anthropology, psychology,...), the knowledge about the laws that dominate their dynamics is still scarce and fragmented. Here, we analyse a wide set of online networks (those financed by advertising) by investigating their value dynamics from several perspectives: the type of law they follow, their geographic scope, and the relationship between economic value and financial value. The results show that the networks are dominated by strongly nonlinear dynamics. The existence of non-linearity is often underestimated in social sciences because it involves contexts that are difficult to deal with, such as the presence of multiple equilibria -- some of which are unstable. Yet, these dynamics must be fully understood and addressed if we aim to understand the recent evolution in the economic, political and social milieus, which are precisely characterised by corner equilibria (e.g., polarization, winner-take-all solutions, increasing inequality) and nonlinear patterns. 

\end{abstract}
\begin{document}

\flushbottom
\maketitle
%
%
\thispagestyle{empty}


\section{Introduction}

Networks have always played a special role for human beings in shaping social relations, forming public opinion, and driving economic equilibria. Trade history since barter has been based on networks of producers, consumers, and intermediaries; the banking sector has developed since the Renaissance on the basis of networks, the currency has emerged as a medium of exchange recognized by networks of individuals, and the entire financial sector has evolved through the development of investor networks.   

More recently, the rise and consolidation of the Internet as an enabling infrastructure has dramatically transformed the ecosystem and its dynamics. The Internet, which is a network of networks, has promoted the creation of comprehensive global connections. Trade has turned into a global network run by intermediaries who leverage the power of networks (of users and producers)\cite{bernard_networks_2018}, finance is based on the matching of networks resulting in very volatile reactions\cite{allen_franklin_networks_2009}, and (crypto)currencies emerge according to a "spontaneous" process driven by network effects\cite{marc_andreessen_why_2019}. It is no coincidence that social relations in the last twenty years have been shaped within a digital sphere dominated by networked platforms\cite{edgerly2009youtube, fuchs2015culture}.   

Unprecedented in the history of capitalism, online networked platforms have in just a few years achieved dominant positions in many digital and traditional markets and have quickly dominated capitalization leaderboards\cite{tirole_competition_2020}. At the end of 2021, the top four companies in the world by market capitalisation were networked platforms (i.e., Apple, Microsoft, Alphabet, and Amazon), which were 6 out of the top 10 (the above four plus Meta and Tencent).

To understand this evolution, therefore, we must understand networks, how they are organised, how they evolve, and what laws they follow. The aim of the article is therefore to examine the dynamic laws that follow new networked platforms also in relation to the laws that followed legacy technologies (TV, peer-to-peer communication,...).

\section{Analysing Networks}

From a general point of view, the concept of network extends to all those structures in which the economic (and social) relations of agents are mutually interdependent (so that "coordination" issues occur). In the classical case, the utility of a service (e.g. a telephone service, a messenger service or a social network) for a user depends on the number of users who have already joined the network. This effect is referred to as "direct network externality". Since platforms are intermediaries linking two (or more) types of agents (e.g. users and advertisers, users and shopkeepers) and thus operate on two- (or more) sided markets these relationships may be indirect. This kind of effects is referred to as "indirect (or cross-side) network externalities". For instance, a search engine exploits the inter-dependency between users and advertisers, by clustering groups and subgroups of users and reselling them on the advertising side (as well as providing customised services); it thus creates and connects networks of agents from seemingly separate domains. In this respect, networks include broadcasting networks (in which a central node transmits a message to peripheral nodes), telephone networks (in which users are linked by peer-to-peer relations), search engines (in which a platform links two or more networks of distinct agents potentially clustered within groups) and, of course, social networks (in which users create group and sub-group relations).  

Notice that such a concept of network is different from what is normally intended in scientific fields like network science \cite{national_research_council_us_network_2005}: in fact, in economics network effects are the phenomena by which the value depends on the number of users without assuming a knowledge of the network topology.

Indeed, networks can be analysed from several aspects. Network analysis investigates the topology of networks, examining their nodes and links, their relation (the density, i.e., the ratio of links to nodes), the directionality of links (either directed or undirected), as well as the formation of clusters. Recent literature on the structure of social relationships within social networks has shown the formation of clusters based on homophily, with little mutual communication between different groups (so-called polarisation). 

The analysis of network topology is fascinating and fundamental to understanding the phenomenon, but it clashes with two major challenges. First, most networks that govern economic and social life (social networks, search engines, e-commerce networks, online trading platforms, etc.) are nowadays private. It is therefore challenging (for reasons ranging from trade secrecy to privacy issues) to acquire enough information to conduct a comprehensive network analysis. Consequently, it is often unfeasible to define general metrics that describe networks comprehensively and concisely (i.e., what statisticians refer to as sufficient statistics\cite{fisher_mathematical_1922}) and can be calculated consistently over time.  

Given the economic and financial prominence of networked platforms in contemporary economies and societies, some metrics have nevertheless been defined. These metrics are primarily based on a single piece of information: the number of participants in the networks. Traditional networks (e.g., TV and radio broadcasting networks) have been evaluated on the basis of contacts reached (e.g., in terms of audience share). More recently, telecommunications companies have been investigated on the basis of the number of users forming their networks (e.g., in terms of mobile SIM cards). Lastly, networked platforms such as social networks are evaluated based on the number of network participants. 

The use of the number of participants to characterise a network has become standard also for reasons concerning economy theory. Since the 1970s\cite{rohlfs_theory_1974}, economic models describing the behaviour of networked markets have included the size of the network within the utility functions of the users. In the presence of (direct and indirect) network effects, as the number of network participants rises, the utility of users increases accordingly, thus producing positive feedback loops and "the winner takes all" market dynamics\cite{shy_economics_2001,tirole_competition_2020}. 

The presence of non-linearities may thus translate into the value of networks. 
In particular, this article aims to test the following hypotheses that appear particularly important for understanding the role that networked platforms have acquired in modern economies and societies: 

\vspace{0.3 cm}

H1: the value of new networked platforms follows dynamics (power-law or exponential) that are faster than those of previous technologies (linear or quadratic).

\vspace{0.3 cm}

H2: the network value dynamics is dependent of the geographical and market context.

\vspace{0.3 cm}

H3: the dynamics of the network's financial value is linked to that of its economic value.

\section{Testing the Dynamics of the Value of Networks}

Over the years, the relationship between network size and network value has been the subject of numerous "laws" that have been proposed by researchers and practitioners. Although these network laws influence important decisions (such as the financial evaluation of companies and related investment decisions), they are neither based on sound theoretical foundations nor on empirical evidence. 

These laws associate the value of the network not only with the number of participants but also with the type of technology. In other words, while the value of the network always depends on the number of participants, its dynamics is related to the generation of the communication technology. 

The value of a broadcasting network (i.e. where a single node communicates unidirectionally to all the others) increases proportionally to the number of users: this relationship is referred to as "Sarnoff's law"\cite{kovarik_revolutions_2015} by David Sarnoff, a titan of the broadcast era radio and TV, who led the Radio Corporation of America (which created NBC) from 1919 until 1970. In Metcalfe’s law\cite{metcalfe_metcalfes_2013} instead, the value of communications networks grows in proportion to the number of possible links among couples of users, i.e. to the square of the number of users on the network. Robert Metcalfe, who was one of the inventors of the Ethernet standard, proposed this law in the 1980s to describe p2p communication networks like fax or phone networks. Finally, in 1999, David Reed\cite{reed_that_1999}, from MIT, proposed, for group-forming networks like instant messaging, that the value of the network increases as $2^N$ (where N is the number of participants) since this number is the number of all possible subgroups that can be formed in a network consisting of N nodes. 

Consequently, the relationship we observe between network value (V) and the number of participants (N) provides insights into the type of connections that characterise the network. 
Sarnoff’s and Metcalfe’s laws have been empirically generalised by Nivi as 
$$V \propto N^\gamma$$;
Sarnoff’s law corresponds to $\gamma = 1$, Metcalfe’s to $\gamma = 2$, while $\gamma > 2$ allows accounting for faster power-law growths. Interestingly, Nivi’s law\cite{nivi_between_2005} can also be justified theoretically with some hypotheses on the way groups form (see Methods).

In this case, the exponent $\gamma$ provides information on the linearity/non-linearity of network value growth with respect to the number of participants. In other words, the estimated value of $\gamma$ indicates whether the value of networked platforms follows a linear dynamic such as that of broadcasting (i.e., $\gamma = 1$), a quadratic pattern as that of peer-to-peer networks ($\gamma = 2$), or a more explosive path ($\gamma > 2$). 

Reed’s law has never been observed in real networks’ data since it can be considered an upper bound to the network value. So, it is possible to generalise Reed’s law (see Methods) to the case of an exponential growth “slower” than the $2^N$ case:
$$V \propto e^{\rho N}$$
corresponding to selecting only a fraction of all possible subgroups. In this case, $\rho$ is the network value’s growth rate bounded by the maximum value  $\rho = \ln 2$ (i.e., $0.69$) corresponding to the original Reed's law.

In the following, we will compare Nivi's growth $V \propto N^\gamma$ to the nested model $V \propto N^2$ corresponding to Metcalfe's growth via the likelihood of the corresponding regressions \cite{li_graduate_2019} . We will then use the information-theoretic approach to model selection \cite{burnham_model_2002}to verify whether the value dynamics of networked platforms follows an exponential (Reed) or power law (Nivi) pattern. 

\section{The Value of Advertising-funded Networks}

Among the most important networks are those funded by advertising. Commercial TV and radio broadcasters, search engines, video sharing platforms, and social networks are two-sided networks that offer free services on the one side (the user side) and make money from selling advertising to advertisers on the other. 

Traditional media networks sell aggregate and undifferentiated contacts. In this case, the value of these networks is related to the number of people who come in contact with the network and thus follows a linear pattern (Sarnoff's law). Online networks, on the other hand, have the ability to sell subgroups (so-called targets) categorised on the basis of specific characteristics (e.g., age, gender, ethnicity, income, location). This results in network values that grow non-linearly with respect to the value of the users (Nivi and/or Reed’s laws). It should be noted that while network effects can benefit (also) users by creating a reinforcement feedback loop that increases the intrinsic non-monetary value of the network and thus expands its user size (so-called bandwagon effect), the non-linearity of the economic value depends on the ability to sell information about users to advertisers and to create different targets and subgroups. In other words, the benefits from network externalities are monetized on the advertising side. 

\subsection{Networked Platforms versus Legacy Technologies}

To verify the dynamics of the value of ad-networks with potentially unlimited connections among participants, we first focus on the US market, where we analyse search engines (i.e., Google), video streaming platforms (i.e., YouTube), and social networking (i.e., Facebook, Instagram, Reddit, Twitter, LinkedIn, Pinterest and Snapchat) over a period of 20 years (i.e. 2002-2021). This choice is consistent with market analysis: “The United States is a relevant geographic market for personal social networking services due to several factors . . . network effects between users are generally stronger between users in the same country, because for most users the vast majority of relevant friends, family, and other personal connections reside in the same country as the user. Accordingly, users in the United States predominantly share content with other users in the United States” (Federal Trade Commission vs. Facebook Inc., Case No.: 1:20-cv-03590-JEB, paragraph169, August 2021). 

To test the dynamics of the network value, we calculate the parameter $\gamma$ on which  we perform a likelihood-ratio test\cite{li_graduate_2019} (also known as Wilks test) respect to Metcalfe's law corresponding to  fixing $\gamma = 2$. In Table \ref{tab:NiviLLratio} we show that the estimated $\gamma$'s are always significantly greater than $2$ (note that the p-values of the Wilks test are all $<0.01$), indicating that the value of networked platforms grows faster than the Metcalfe's law (i.e., legacy technologies). In addition, the values of the coefficients of determination $R^2\sim 1$ indicate a good accordance with data. Accordingly, we observe that Nivi's model has a better predictive performance than Metcalfe's (Table \ref{tab:PredPerfMNR}).

\subsection{Model Selection: power law versus exponential}

\begin{figure}[!ht]
    \centering
    \includegraphics[width=1.0\columnwidth]{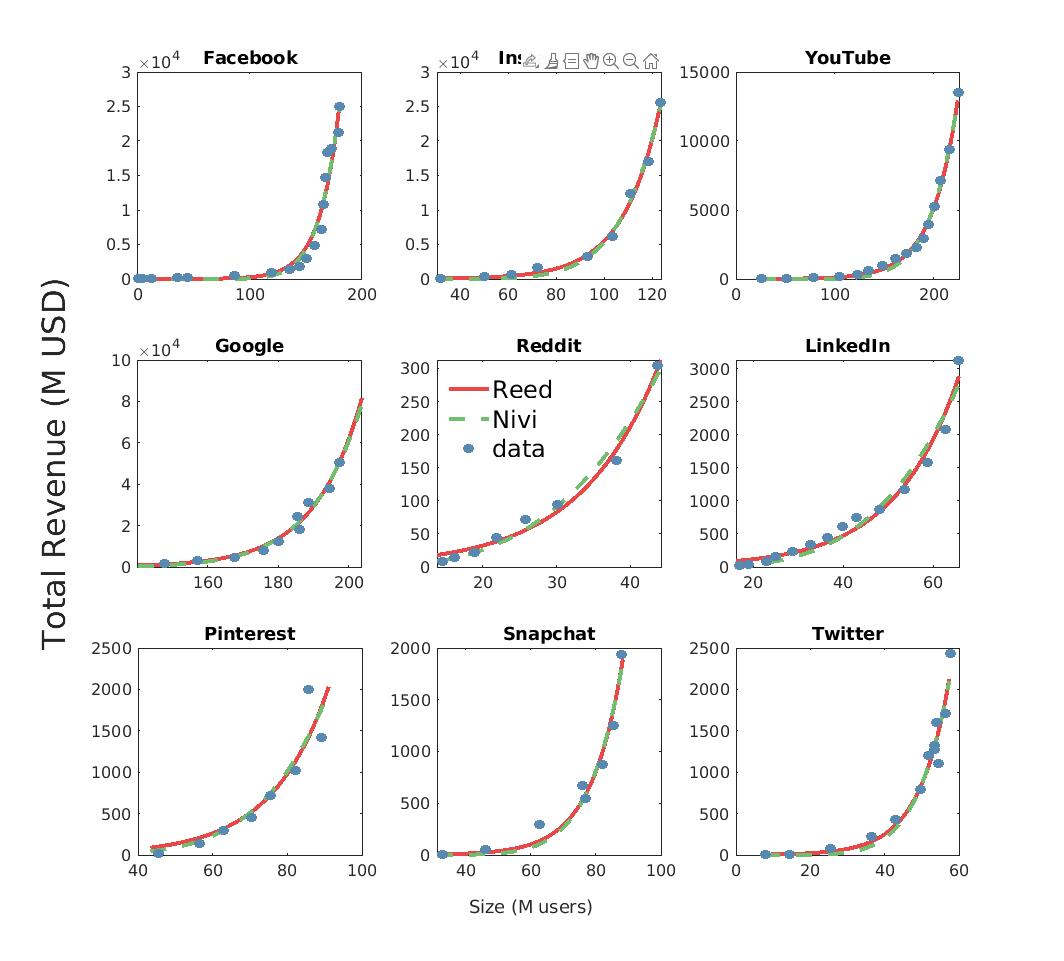}
\caption{ \textbf{Network Value for online services in the US market.} The advertising revenues for several US network companies grow faster than quadratically and are better fitted either by the generalised Reed’s law (exponential growth) or by Nivi’s law (power law growth), with exponents varying from $\gamma \sim 3$ to $\gamma \sim 14$ (see Table \ref{tab:NiviLLratio}).}
\label{fig:ReedNiviUS}
\end{figure}

\begin{figure}[!ht]
    \centering
    \includegraphics[width=1.0\columnwidth]{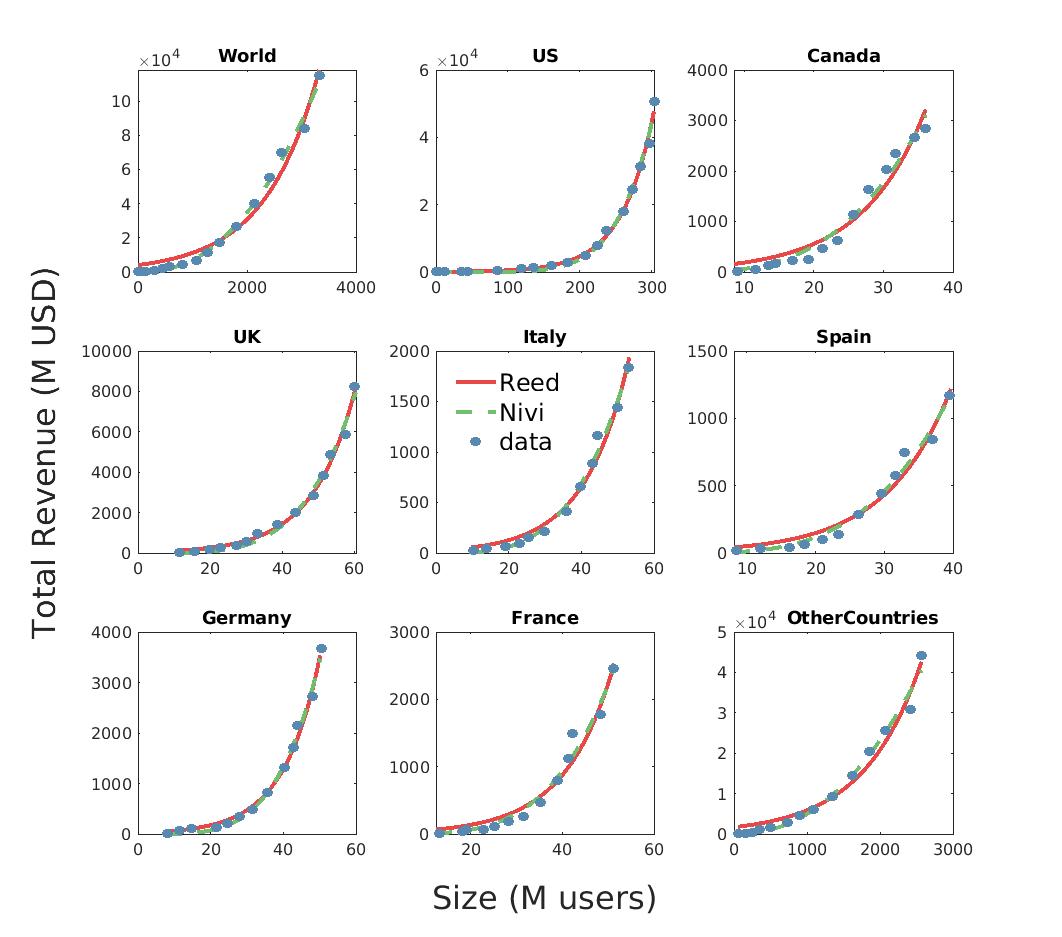}
    \caption{\textbf{Network Value for the geografical areas of META.} The advertising revenues grow faster than quadratically and are better fitted either by the generalised Reed’s law (exponential growth) or by Nivi’s law (power law growth), with exponents varying from $\gamma \sim 2.2$ to $\gamma \sim 6$ (see Table \ref{tab:NiviLLratio}).}
    \label{fig:ReedNiviMeta}
\end{figure}

In Fig. \ref{fig:ReedNiviUS} and \ref{fig:ReedNiviMeta} we show the accordance of the data with the fits of Reed's (exponential) and Nivi's (power law) models both for multiple platforms in the US market and for the different geographic market areas (of Meta, i.e., Facebook + Instagram). Neither visual inspection nor the $R^2$ values (see also Sec.\ref{sec:SI}) allow to select a better model. Since Reed's and Nivi's laws cannot be defined as nested statistical models, standard log-likelihood ratio test cannot be applied \cite{burnham_model_2002}. Thus, we have employed the information-theoretic approach to model selection \cite{burnham_model_2002} to discriminate among the two models (see also Sec.\ref{sec:SI}). As shown in Table \ref{tab:AkaikeNR}, the data are not conclusive (apart in the case of META's aggregated World data) in discriminating whether Reed's or Nivi's is the best model, indicating some heterogeneity linked to the geographical context (it generally follows an exponential pattern in the US, and a power law elsewhere). Accordingly, we observe that - apart for META's aggregated World data, the predictive performance of both models are comparable (see Table \ref{tab:PredPerfMNR}).

Finally, it should be noted that, as expected (see Methods), $\rho$ is always significantly less than $0.69$ (see above), displaying predictable variability depending on services (it is greater in social networks than in search engines and video sharing), and geographical context (it is greater in national markets than in transnational areas such as the World or Other countries).

\subsection{Determinants of Value Network Dynamics}

\begin{figure}[!ht]
    \centering
    \includegraphics[width=1.0\columnwidth]{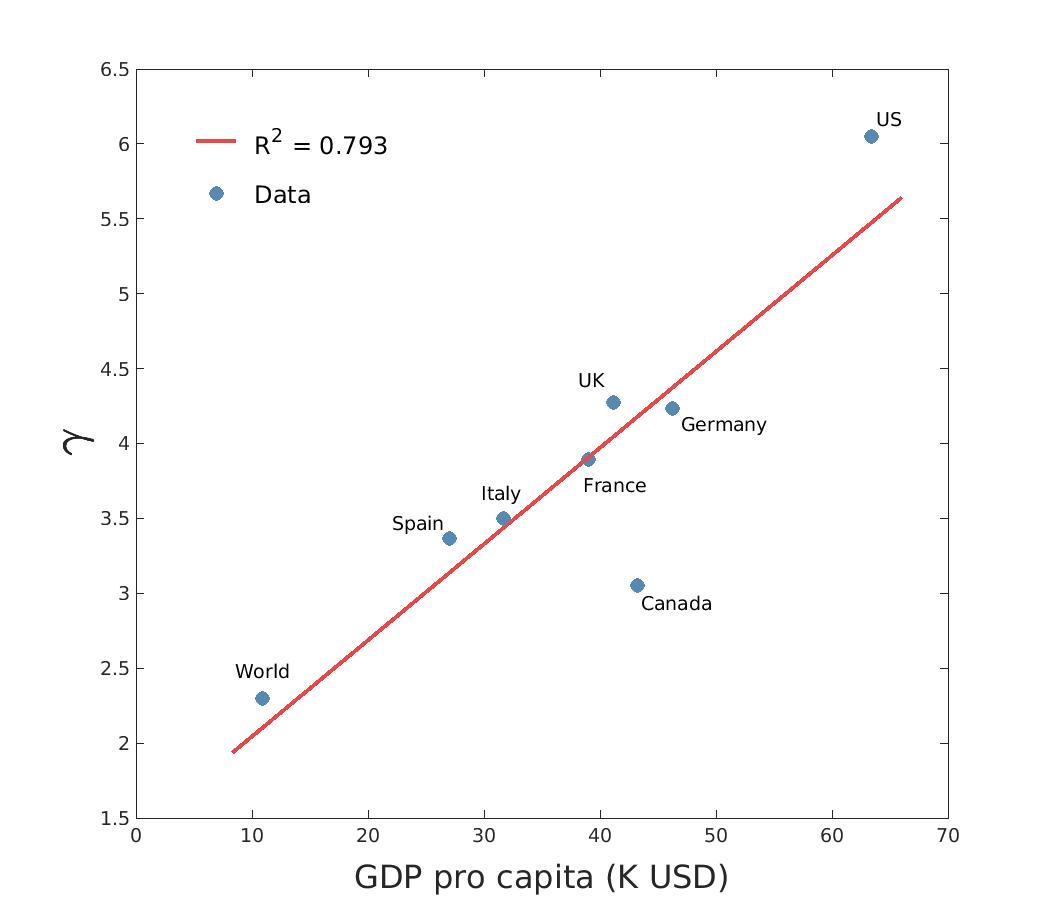}
    \caption{Meta data: relation between Nivi's exponent $\gamma$ and GDP pro capita (in thousands of USD) in different geographical areas.
    }
    \label{fig:gammaVSgdpMeta}
\end{figure}

To check for heterogeneity in the network growth across different geographical areas, we investigate the data for the global platform Meta (see Fig. \ref{fig:ReedNiviMeta}). A shown in Table \ref{tab:NiviLLratio}, we observe again that: i) the estimated $\gamma$'s are significantly greater than $2$; ii) there is a good accordance with data as indicated by the values of the coefficients of determination $R^2\sim 1$; iii) We observe a strong correlation of $\gamma$ with the per-capita income of different countries ($R^2 = 0.79$, see Fig. \ref{fig:gammaVSgdpMeta}), hinting that $\gamma$ summarises the platform's ability to economically leverage its network of users. 
A possible explanation of such effect is that where per-capita income is higher, users' willingness to pay for consumer goods is higher, thus increasing the opportunity to economically exploit, through the sale of advertising, the network 

Note also that the value of $\gamma$ is highest in the case of dominant platforms in each service (Google for search engines, Facebook for social networking and YouTube for video sharing platforms). With specific reference to social networks, it should be also noted that the values of $\gamma$ for LinkedIn and Reddit is lower probably because both networks rely, in addition to advertising, on premium membership revenues. 


\subsection{Economic and Financial Value}

To investigate the relation between the network value and is financial value using as a measure its market capitalization, we analyse publicly traded network companies for the World market; i.e. a search engine (Google) and three social networking platforms (Meta, Twitter and Snapchat). It should be noted that the financial value reflects the market's expected value of the present and future discounted profits of the networked platform. 

\begin{figure}[!ht]
    \centering
    \includegraphics[width=1.0\columnwidth]{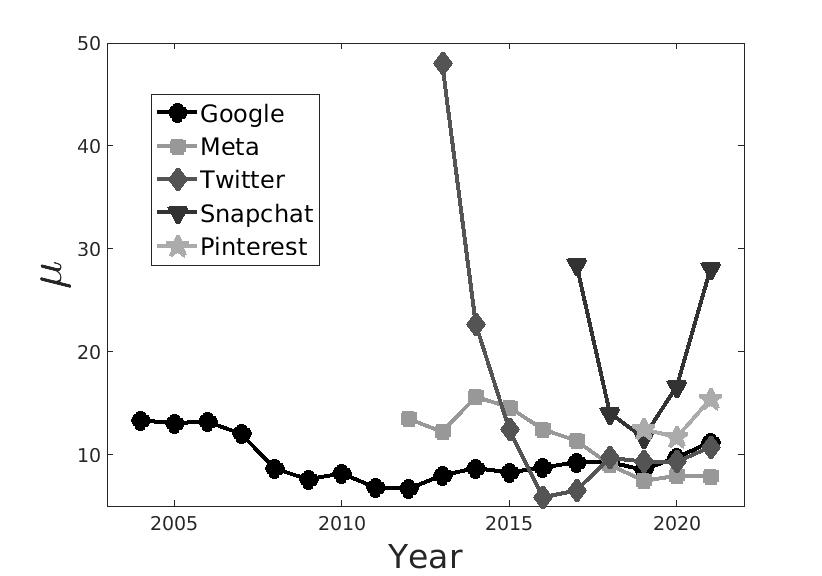}
\caption{\textbf{Market capitalization over Network value ratio.} The yearly behaviour of the multiplier $\mu$, corresponding to the ratio between the market capitalization and the total return on advertising, shows convergence to a plateau around $\mu \sim 10$ (but for Snapchat -- which however is converging to the plateau in 2022).}

    \label{fig:WORLDmultiplier}
\end{figure}

To capture the relation among the network value $V$ and its financial value $F$, we study the ratio $\mu=F/V$ among the two; such ratio is a multiplier of the economic value of the network. A constant ratio over time would indicate that the market prizes the company according to its sales. In Fig. \ref{fig:WORLDmultiplier} we show the time evolution of the ratio $\mu$ among market capitalization and advertising revenues. We notice that for Google, Meta and Twitter such multiplier stabilises around $\mu \sim 10$; a similar value is attained by considering also Pinterest data. On the other hand, Snapchat data are anomalous in such regard: in fact, after reaching a value $\mu \sim 10$ from 2017 to 2019, it grows up to an anomalous value thrice larger in 2021. However, it should be noted that in accordance with our observations, Snapchat's market value dropped significantly in 2022, such that it fell between January and June by nearly 70\%, yielding an estimated multiplier $\mu \sim 9$ consistent with the plateau values of Fig. \ref{fig:WORLDmultiplier}.

In sum, data seem to show that in the medium term $ F = \mu V$ along all  platforms, indicating that the financial market prizes these companies proportionally to their economic network value, so that in equilibrium (regardless of speculative perturbations) also the financial value follows "explosive" growth laws.

\section{Discussion}

Economies and societies are based on networks of agents and citizens, and their governance has dramatically changed in recent decades with the establishment of the Internet as an enabling infrastructure and the resulting shift in socio-economic dynamics toward nonlinear patterns. Yet surprisingly, knowledge of these dynamics is still scarce and based on scientifically unsound "laws" and scant empirical evidence.  

Here we developed a comprehensive model and tested the dynamics of the (economic and financial) value of online ad-funded networks. The results show a strongly nonlinear trend, especially when focusing on a specific geographical area.
In particular we have found that:
\begin{itemize}
\item the economic value of networked platforms (i.e., search engines, social networks, and video sharing platforms) follows a dynamic that is faster than that characterizing previous generation technologies (linear, i.e., Sarnoff law, for broadcasting, and quadratic, Metcalfe law, for peer-to-peer networks);
\item the estimated models are in most cases not conclusive with respect to the type of nonlinear dynamics (i.e., whether power law or exponential); however, they show heterogeneity depending on the type of service and the geographical context;
\item in this regard, the dynamics appear to depend on factors such as the market power gained by the networked platform and the average willingness to pay of local users; 
\item finally, the financial value of networked platforms, and its dynamics, is intertwined, at least in the medium term, with the economic value. Put another way, the explosive dynamic found at the economic level  appears to reverberate on the value that financial markets assign to these platforms.      
\end{itemize}

Despite the economic and social importance of the phenomena analyzed, to the authors' knowledge, this is the first analysis to investigate the dynamics just described. This article is therefore a first step that foreshadows a variety of interesting future research developments. 
First, the availability of more granular data -- at network, geographical and time levels -- would give the possibility of more observations and thus even more comprehensive results. In fact, while out tests indicate that network platforms grow faster than legacy technologies, the present evidence is not enough to discriminate among Nivi's or generalised Reed. Second, by gathering historical data on legacy technologies (broadcasting networks, peer-to-peer networks), different dynamical patterns could be compared in a robust way. Finally, public or otherwise available data that would allow to estimate topological network statistics (like data on the links among the users) would clarify further aspects of the analyzed dynamics that go beyond economic concepts related to economic network effects.

\section{METHODS}

\subsection{Data Sources and analysis}

Data on users and advertising revenues are, where available, from eMarketer. Data can be requested via access to the \href{https://www.insiderintelligence.com/}{Insider Intelligence} database. 
These data cover both the US and global markets (and for Meta also Canada, France, Germany, Italy Spain, and UK). For the early years of operation of "older" networked platforms (before 2008) these data are often not available. In that case, they have been supplemented by other sources such as companies' financial statements (e.g., Google, Alphabet), other official communications to the financial community (e.g. IPOs), and \href{https://www.statista.com/}{Statista}. Data on Google users are elaborations from \href{https://www.comscore.com/}{comScore} data. Data on countries' GDP per-capita are from the \href{https://data.worldbank.org/}{World Bank} and pertain to the year 2021. Finally, data on the market capitalization of companies are average annual values computed from daily data from \href{https://www.bloomberg.com/}{Bloomberg} (as of the day of IPO).  

To determine which law fits better our data on the network values, we apply non-linear least square minimization of the following functional forms
\[ 
\begin{array}{ll}
V =  a_M N^2 & \textrm{Metcalfe} \\
V =  a_N N^\gamma & \textrm{Nivi}\\
V = a_R e^{\rho N} & \textrm{Reed}
\end{array}
\]
that capture the scaling of network values respectively for Metcalfe's, Nivi's and generalised Reed's law. 

Since Metcalfe can be seen as a nested model of Nivi (its parameter are a subset of Nivi's since Metcalfe can be obtained by setting $\gamma=2$), log-likelihood ratio tests \cite{li_graduate_2019}  allow to perform model selection among the twos along the different datasets. On the other hand, since Reed's and Nivi's laws cannot be defined as nested statistical models, we have applied the information-theoretic approach to model selection \cite{burnham_model_2002} to discriminate among the two models along the different datasets by confronting Akaike's $\Delta$'s (see also Sec.\ref{sec:SI}). 

To test model's predictive performance, we first calculate via a leave-one-out procedure the mean squared prediction error (MSPE) \cite{gareth2013introduction}. To compare models, we look at the ratio of the MSEs respect to the one for Metcalfe's (see Table \ref{tab:PredPerfMNR}); if a model performs better than Metcalfe's, its ratio will be $<1$, while the ratio will be $>1$ for worse prediction performing models. As expected, Nivi's and Reed's model show significantly lower MSPE ratios but for the cases of META's data for World and Other countries, which both have an exponent $\gamma \sim 2$. Further details can be found in Sec.\ref{sec:SI}.

In the following, we derive the generalised Reed’s law under the minimal assumption that new users contribute to the formation of new groups at a reduced rate with respect to the original Reed’s law. We then derive the empirical Nivi’s law under the assumption that the rate of new groups’ formation is decreasing with the number of users. 

\subsection{Derivation of the network value’s growth laws}

The advertising value of a network is related to the possibility of selling groups of users for targeted advertising; however, Reed's intuition that network value should be linked to the number $2^N$ of all possible subgroups has been criticised for having a too sharp increase of the network value\cite{odlyzko_refutation_2005} never comforted by observational data\cite{zhang_tencent_2015}. Moreover, Reed’s Law cannot be economically correct: predicting that a network of $1010$ users would have a value $2^{10} \sim 1000$ larger than a network of 1000 users clashes with our expectations (and experience) about network values. It is in fact unrealistic to expect that each of the possible $2^N$ groups can be sold to a different advertiser. Is it possible to adapt Reed's core insight that there is value in group formation without violating common sense and economic data?

To develop simple models that restrict the possible groups to the "marketable" ones (i.e. the ones that can be sold), let's observe that the number $\Omega(n)$ of "marketable" subgroups of n users grows as $\Omega(n+1) = \Omega(n) + 	\Delta \Omega(n)$ where $\Delta \Omega(n)$ is the number of new groups that can be formed by adding a new user. In the original Reed's approach, each new user allows creating of  $\Omega(n)$ new groups, leading to $2^N$ groups for N users. However, in a real market, not all the new groups are saleable; thus we can suppose that a new individual has only a probability $\rho$ of determining a new "marketable" group from an old one,  i.e. $\Delta \Omega(n)  = \rho \Omega(n)$. Lets notice that, as soon as $n>> 1$, it is convenient to express the growth in differential form, i.e. $ \partial_n \Omega(n) =  \Delta \Omega(n) $;  the solution of such a growth
$$\partial_n \Omega(n) = \rho \Omega(n)$$recovers exactly what we have called the generalised Reed's growth
$$\Omega(n) \propto  e^{\rho n}$$
. The observed estimated values of $\rho$ are extremely small ($\sim 10^{-9}$, see Table \ref{tab:AkaikeNR} and Sec.\ref{sec:SI}), indicating that the probability that a user determines a new group is extremely small.

On the other hand, it is reasonable to assume  that while at the beginning of the network formation it is very easy to create new "marketable" groups, it should become increasingly difficult to find new "marketable" groups when $n$ grows. Accordingly, we should expect a growth law 	$\Delta \Omega(n) = p(n)	\Omega(n)$ where the probability $p(n)$ of creating new "marketable" groups is now decreasing with $n$. An instructive and solvable case is to assume a simple Zipf's law $p(n)\propto 1/n$; such a law was empirically found in quantitative bibliographic studies indicating that the probability of finding "new" or "interesting" papers decreases approximately like the inverse of the number of analysed papers. Thus, assuming $p(n) = \gamma/n$, we have a growth law 
$$\partial_n \Omega(n) = \frac{\gamma}{n} \Omega(n)$$
whose solution
$$\Omega(n) \propto  n^\gamma$$
recovers the empirical Nivi's law.


\newpage

\begin{table}[!ht]
\centering
\begin{tabular}{ |l|ccc| }
\hline 
Network (US) & $\gamma-2$ & $p$-value & $R^2$\\ 
\hline 
Facebook & 7.7 & 3.2e-10 & 0.958 \\ 
Instagram & 5.3 & 1.3e-08 & 0.992 \\ 
YouTube & 5.9 & 2.2e-16 & 0.994 \\ 
Google & 12.1 & 1.2e-08 & 0.974 \\ 
Reddit & 1.1 & 4.9e-4 & 0.980 \\ 
LinkedIn & 1.6 & 2.6e-05 & 0.963 \\ 
Pinterest & 3.1 & 4.8e-3 & 0.863 \\ 
Snapchat & 6.4 & 6.4e-05 & 0.955 \\ 
Twitter & 4.5 & 8.3e-06 & 0.953 \\ 
\hline 
\end{tabular}
\hspace{1cm}
\begin{tabular}{ |l|ccc| }
\hline 
Meta & $\gamma-2$ & $p$-value & $R^2$\\ 
\hline 
World & 0.3 & 2.0e-4 & 0.995 \\ 
US & 4.0 & 2.9e-14 & 0.991 \\ 
Canada & 1.0 & 1.3e-05 & 0.977 \\ 
UK & 2.3 & 3.8e-09 & 0.988 \\ 
Italy & 1.5 & 4.9e-08 & 0.991 \\ 
Spain & 1.4 & 2.1e-06 & 0.982 \\ 
Germany & 2.2 & 8.1e-11 & 0.994 \\ 
France & 1.9 & 3.0e-07 & 0.977 \\ 
OtherCountries & 0.2 & 9.9e-2 & 0.985 \\
\hline 
\end{tabular}

\caption{ \textbf{Test for Nivi's exponents.} 
Left table: results for US networked platforms. Right table: results for Meta's network by geographical area. In each table we report the difference $\gamma-2$ between the estimated value of Nivi's exponent and Metcalfe's; the $p$-value obtained by the Wilk's test \cite{li_graduate_2019}; and the coefficient of determination $R^2$. The $\gamma-2$ value provides an estimation of the deviation of the Nivi's growth rate from Metcalfe's and the $p$-value is calculated  with the null hypothesis that $\gamma = 2$  (i.e., Metcalfe law). 
\label{tab:NiviLLratio}}
\end{table}

\newpage

\begin{table}[!ht]
\centering
\begin{tabular}{ |l|ccc|ccc| }
\hline 
\multicolumn{1}{|l|}{Network (US)}&\multicolumn{3}{|c|}{Nivi}&\multicolumn{3}{|c|}{Reed}\\~ & $\gamma$ & $R^2$ & $\Delta$ & $\rho$ & $R^2$ & $\Delta$ \\ 
\hline 
Facebook & 9.7 & 0.958 & 0.0 & 5.7e-2 & 0.954 & 1.6 \\ 
Instagram & 7.3 & 0.992 & 2.2 & 6.5e-2 & 0.994 & 0.0 \\ 
YouTube & 7.9 & 0.994 & 5.4 & 3.8e-2 & 0.996 & 0.0 \\ 
Google & 14 & 0.974 & 0.0 & 7.5e-2 & 0.973 & 0.3 \\ 
Reddit & 3.1 & 0.980 & 3.3 & 9.4e-2 & 0.987 & 0.0 \\ 
LinkedIn & 3.6 & 0.963 & 6.7 & 7.0e-2 & 0.977 & 0.0 \\ 
Pinterest & 5.1 & 0.863 & 0.0 & 6.6e-2 & 0.854 & 0.6 \\ 
Snapchat & 8.4 & 0.955 & 1.5 & 1.0e-1 & 0.963 & 0.0 \\ 
Twitter & 6.5 & 0.953 & 1.4 & 1.2e-1 & 0.958 & 0.0 \\ 
\hline 
\end{tabular}

\vspace{1cm}

\begin{tabular}{ |l|ccc|ccc| }
\hline 
\multicolumn{1}{|l|}{Meta}&\multicolumn{3}{|c|}{Nivi}&\multicolumn{3}{|c|}{Reed}\\
~ & $\gamma$ & $R^2$ & $\Delta$ & $\rho$ & $R^2$ & $\Delta$ \\
\hline 
World & 2.3 & 0.995 & 0.0 & 0.1e-2 & 0.977 & 28.6 \\ 
US & 6.0 & 0.991 & 4.0 & 2.3e-2 & 0.993 & 0.0 \\ 
Canada & 3.0 & 0.977 & 0.0 & 1.1e-1 & 0.951 & 10.6 \\ 
UK & 4.3 & 0.988 & 6.5 & 8.5e-2 & 0.992 & 0.0 \\ 
Italy & 3.5 & 0.991 & 0.0 & 8.2e-2 & 0.983 & 8.0 \\ 
Spain & 3.4 & 0.982 & 0.0 & 1.1e-1 & 0.967 & 7.3 \\ 
Germany & 4.2 & 0.994 & 0.0 & 1.0e-1 & 0.994 & 1.0 \\ 
France & 3.9 & 0.977 & 0.0 & 9.4e-2 & 0.963 & 6.0 \\ 
OtherCountries & 2.2 & 0.985 & 0.0 & 1.3e-3 & 0.975 & 6.9 \\ 
\hline 
\end{tabular}

\caption{ \textbf{Akaike test for Reed's and Nivi's law.} 
The Akaike test allows to discriminate among two statistical models that are not nested by calculating the difference among the value $\Delta$ of their Akaike information criteria. The empirical level of support of a model from the data is substantial for $\Delta \sim 0-2$, considerably less for $\Delta \sim 4-7$ and essentially none for $\Delta >10$ \cite{burnham_model_2002}. We notice that, the coefficients of determination $R^2$ are similar for the two models and that Akaike's $\Delta$s seldom rule out completely the possibility that the "least good" model is also supported by the data.
\label{tab:AkaikeNR}}
\end{table}

\newpage

\begin{table}[!ht]
\centering

\begin{tabular}{ |l|ccc| }
\hline 
Network (US) & Metcalfe & Nivi & Reed\\ 
\hline 
Facebook & 1.00 & 0.15 & 0.16 \\ 
Instagram & 1.00 & 0.06 & 0.04 \\ 
YouTube & 1.00 & 0.01 & 0.01 \\ 
Google & 1.00 & 0.05 & 0.05 \\ 
Reddit & 1.00 & 0.51 & 0.17 \\ 
LinkedIn & 1.00 & 0.53 & 0.32 \\ 
Pinterest & 1.00 & 1.01 & 1.18 \\ 
Snapchat & 1.00 & 0.35 & 0.29 \\ 
Twitter & 1.00 & 0.35 & 0.30 \\ 
\hline 
\end{tabular}
\hspace{1cm}
\begin{tabular}{ |l|ccc| }
\hline 
Meta & Metcalfe & Nivi & Reed\\ 
\hline 
World & 1.00 & 0.66 & 1.92 \\ 
US & 1.00 & 0.09 & 0.07 \\ 
Canada & 1.00 & 0.41 & 0.89 \\ 
UK & 1.00 & 0.18 & 0.09 \\ 
Italy & 1.00 & 0.10 & 0.20 \\ 
Spain & 1.00 & 0.18 & 0.31 \\ 
Germany & 1.00 & 0.06 & 0.04 \\ 
France & 1.00 & 0.15 & 0.23 \\ 
OtherCountries & 1.00 & 1.24 & 1.09 \\ 
\hline 
\end{tabular}

\caption{ \textbf{Predictive performance.} 
To compare the predictive performance of the different models, we evaluate the ratio among the model's leave-one-out mean squared prediction error respect to one of the Metcalfe's model. Thus, a ratio $<1$ indicates a better prediction performance while a ratio $>1$ indicates a worse performance. 
Left table: results for US networked platforms. Right table: results for Meta's network by geographical area. 
Notice that larger ratios of Reed's model for META's aggregated world data and for META's other countries data correspond to the cases where $\gamma \sim 2$, i.e the Nivi's model estimate is close to Metcalfe.
\label{tab:PredPerfMNR}}
\end{table}

\newpage

\newpage

\newpage

\newpage

\section{Supplementary Information \label{sec:SI}}

\vspace{1cm}

\section*{S1 Network Value growths}

Let's consider a system of $N$ users. When linking together such users, we can define different possible “network values” $V$ \cite{tongia_dark_2010}:
\begin{itemize}
\item{Sarnoff :} $V\propto N$, corresponding to the possibility of broadcasting a message to all the $N$ users \cite{kovarik_revolutions_2015}. 
\item{Odlyzko-Tilly :} $V \propto N \ln N$, corresponding to assuming that each users has a utility proportional to $\ln N$ (i.e. the probability that his $k^{th}$ connection is useful to him is proportional to $1/k$) \cite{briscoe_metcalfes_2006}. 
\item{Metcalfe :} $V \propto N^{2}$, corresponding to giving a constant value to all the possible links among the $N$ units \cite{metcalfe_metcalfes_2013}. 
\item{Nivi :} $V \propto N^{\gamma}$, a phenomenological power-law growth introduced in \cite{nivi_between_2005}. 
\item{Reed :} $V \propto 2^N$, corresponding to giving a constant value to all the possible subsets of $N$ units \cite{reed_that_1999}. 
\end{itemize}

Thus, Sarnoff \cite{kovarik_revolutions_2015} relates the value to the possibility of reaching all the users of the network at once, like in the old broadcasting media (radio, television).
On the other hand, Metcalfe relates the value to the number of possible connections among users of the system \cite{metcalfe_metcalfes_2013}; it was originally presented in terms of "compatible communicating devices" (i.e. the concept is born after the introduction of telephone networks) and then applied to describe Ethernet connections. 
Since it is unreasonable to assume that all the possible connections are useful to each user (i.e the number of "useful" connection is limited at least by bounded rationality constraints such as Dunbar's number \cite{dunbar_neocortex_1992}), Metcalfe himself proposed that the number of useful connections could eventually saturate \cite{metcalfe_metcalfes_2013}, leading for large $N$ to a linear growth (like Sarnoff) with a large prefactor. 
On the same footing, inspired by search-engine, Odlyzko-Tilly\cite{odlyzko_refutation_2005} models the value per user for finding useful contacts in the network with a Zipf statistics; however, in this case, Odlyzko-Tilly should be modified as $V \propto N \ln M$ where $M$ is the number of objects that users can retrieve.
Reed related the value of a network to the possibility of forming groups (like on social media). 
Common-sense suggest that Reed’s laws, as stated, is incorrect: since every new users doubles the value of a network, no matter how small the prefactor, it would quickly reach the value of the world economy. Notice that, to explain network value growths stronger than Metcalfe but not as steep as Reed's $2^N$, Nivi proposed empirically to use power laws \cite{nivi_between_2005}.

To determine which law fits better our data on the network values, we apply non-linear least square minimization of the following functional forms
\begin{equation*}
\begin{array}{lll}
V =  a  N & ~ & \textrm{Sarnoff} \\
V =  a \, N \ln N & ~ & \textrm{Odlyzko-Tilly} \\
V =  a N^2 & ~ & \textrm{Metcalfe} \\
V =  a N^\gamma & ~ & \textrm{Nivi} \\
V = a e^{\rho N}& ~ & \textrm{generalised Reed} \\
\end{array}
\end{equation*}
that capture the scaling of network values respectively for Sarnoff's, Otziko-Tilly's, Metcalfe's, Nivi's and generalised Reed's model. In Table \ref{tab:R2US} we report the coefficients of determination $R^2$ of the regression for the US dataset to the various growth laws, while in Table \ref{tab:R2MetaAll} we report the $R^2$ for the regressions on Meta's revenues per geographical area.

\section*{S2 Model selection and prediction power}

\subsection*{Nested Models selection}

The goodness of fit of two competing models can be compared based on the ratio of their likelihoods or, equivalently, on the difference of their log-likelihood. The likelihood-ratio test, also known as Wilks test, requires that the models to compare are nested, i.e. that one model can be reduced into the other by fixing some parameters. Regression routines of modern languages for data analysis can furnish as output robust log-likelihood estimates of the regression model. By indicating with $\mathcal{L}_X$ the likelihood of the best estimate of model $X$, and by $\mathcal{L}_Y$ the likelihood of the best estimate of model $Y$ nested in $X$, the ratio
\[ \lambda =-2\ln \frac{\mathcal{L}_Y}{\mathcal{L}_X}\]under certain conditions \cite{li_graduate_2019} is distributed as $\chi^2_f$ where the number of degrees of freedom $f=x-y$ with $x$, $y$ being the number of parameters of $X$, $Y$. Thus, the $p$-value corresponding to the calculated $\lambda$ allows to have an estimate of the likelihood that the our data are described just from the "restricted" model $X$. For the data analysed in the main paper, in the case of Nivi's and Metcalfe's models (the former can be reduced to the latter by costraining $\gamma$ to be equal to $2$) we find by the log-likelihood ratio test that in almost all the cases the null hypothesis $\gamma=2$ corresponds to $p$-values less than $10^{-2}$, i.e. it is extremely unlikely. 

\subsection*{Information-theoretic approach to model selection}
Like in log-likelihood ratio tests, model selection based on some form of statistical null hypothesis testing requires that one model is reducible to the other by constraining the parameter space. On the other hand, model selection based on information theory represents a quite different approach in the statistical sciences: while null hypothesis approaches are more suitable for classic experiments (control/treatment with randomization and replication), the information-theoretic approach is more suitable for the analysis of observational studies' data \cite{burnham_model_2002}. 
Starting from the concept of Kullback-Leibler distance $I(A,T)$ that measures the information lost when using a model $A$ to approximate the real behavior $T$, Akaike \cite{akaike1973information,akaike1983information} showed  that “an information criterion” (AIC) \[AIC = -2\log\mathcal{L} -2\,K\]
estimates the expected, relative distance between the model fitted to the data and the unknown true process that actually generated the observations; here the bias-correction term $K$ indicates the number of estimable parameters. In our analysis, since when the number of observations $n$ is low (say $n/K<40$) ,  we use the small sample AIC \cite{sugiura1978further} \[AIC_c = -2\log\mathcal{L} +\frac{2\,K\,(K+1)}{n-K-1}\]
that uses a modified biased correction term.
To proceed to model selection, we compute  the $AIC$ differences \[ \Delta_i = AIC_i - AIC_{min}\]
over all the candidate models. Such differences estimate the \textit{relative} expected K-L distances between the model an the "true" mechanism. The $\Delta_i$  allow to compare and rank candidate models; obviously, the best model has $\Delta_i=0$. When the number of models to compare is small, some rough rules of thumb \cite{burnham_model_2002} are available:
\begin{table}[!htbp]
\begin{tabular}{ccc}
$\Delta_i$ & ~ & Level of Empirical Support of Model $i$  \\
\hline
$0-2$ & ~ & Substantial \\
$4-7$ & ~ & Considerably less \\
$>10$ & ~ & Essentially none \\
\end{tabular}
\end{table}

\noindent Notice that since the likelihood of a model given the data is $\propto \exp(-\Delta_i /2)$, Akaike advocated the use of Akaike's weights \[ w_i = -\frac{e^\frac{\Delta_i}{2}}{\sum_i e^\frac{\Delta_i}{2}}\]. The Akaike weights not only provide an effective way to scale and interpret the  $\Delta_i$ values, but also allow to calculate the evidence ratios $w_i/w_j$ corresponding to the relative likelihood of occurrence of the model pairs $i,j$ \cite{burnham_model_2002}.

\subsection*{ Models' predictive performance}

To estimate the predictive performance of a model, an useful quantity is the mean squared prediction error measuring the accuracy of a model's prediction on a subset of data when the model's parameters are regressed to the remaining data. In particular, since our data sets contain a limited number of observations, we will use just the mean squared prediction error calculated via a leave-one-out (LOO) procedure, i.e. predicting the $i^{th}$ observation by fitting the remaining $n-1$.

In our case, for each model we calculate the $MSPE$  \cite{gareth2013introduction} \[MSPE=\frac{1}{n}\sum_i (V_i - \hat{V_i})^2 \]where $n$ is the size of the dataset, $V_i$ is the $i^{th}$ observed network value (aka its advertising revenues),  $\hat{V_i}$ is its estimate by fitting the model on all observation but the $i^{th}$. Given a dataset, a model will have a better predictive performance (in our case, on single data) if it has a lower $MSPE$. Thus, to compare the predictive performance of the different models, we evaluate the ratio among the model’s  $MSPE$ respect to Metcalfe’s model $MSPE$. Thus, a ratio < 1 indicates a better prediction performance respect Metcalfe's model, while a ratio > 1 indicates a worse performance.

\newpage
\begin{table}[!ht]
\centering
\begin{tabular}{ |l|ccccc| }
\hline 
(US) & Reed & Nivi & Metcalfe & Odlyzko-Tilly & Sarnoff \\ 
\hline 
Facebook & 0.954 & 0.958 & 0.622 & 0.506 & 0.467 \\ 
Instagram & 0.994 & 0.992 & 0.710 & 0.532 & 0.461 \\ 
YouTube & 0.996 & 0.994 & 0.625 & 0.454 & 0.401 \\ 
Google & 0.973 & 0.974 & 0.333 & 0.209 & 0.177 \\ 
Reddit & 0.987 & 0.980 & 0.909 & 0.726 & 0.600 \\ 
LinkedIn & 0.977 & 0.963 & 0.869 & 0.691 & 0.588 \\ 
Pinterest & 0.854 & 0.863 & 0.631 & 0.450 & 0.379 \\ 
Snapchat & 0.963 & 0.955 & 0.666 & 0.512 & 0.444 \\ 
Twitter & 0.958 & 0.953 & 0.785 & 0.669 & 0.602 \\  
\hline 
\end{tabular}
\caption{\label{tab:R2US}US data: $R^2$ values for Reed's, Nivi's, Metcalfe, Sarnoff's and Odlyzko-Tilly's laws.
}
\end{table}

\newpage
\begin{table}[!ht]
\centering
\begin{tabular}{ |l|ccccc| }
\hline 
META & Reed & Nivi & Metcalfe & Odlyzko-Tilly & Sarnoff \\ 
\hline 
World & 0.977 & 0.995 & 0.990 & 0.875 & 0.836 \\ 
US & 0.993 & 0.991 & 0.785 & 0.620 & 0.567 \\ 
Canada & 0.951 & 0.977 & 0.910 & 0.751 & 0.628 \\ 
UK & 0.993 & 0.988 & 0.857 & 0.687 & 0.588 \\ 
Italy & 0.983 & 0.992 & 0.898 & 0.733 & 0.629 \\ 
Spain & 0.968 & 0.982 & 0.885 & 0.716 & 0.600 \\ 
Germany & 0.994 & 0.994 & 0.854 & 0.691 & 0.594 \\ 
France & 0.963 & 0.977 & 0.826 & 0.633 & 0.525 \\ 
OtherCountries & 0.975 & 0.985 & 0.982 & 0.872 & 0.830 \\ 
\hline 
\end{tabular}
\caption{\label{tab:R2MetaAll}Meta data: $R^2$ values for Reed's, Nivi's, Metcalfe, Sarnoff's and Odlyzko-Tilly's laws over different countries.
}
\end{table}

\end{document}